\documentclass[a4paper,twoside]{article}
\newcommand{\affil}[1]{$^{\rm #1}$}
\usepackage[authoryear]{natbib}
\bibpunct{(}{)}{;}{a}{}{,}
\usepackage{graphicx}
\date{} %Please leave the date blank
\title{\large\bf\flushleft Galactic chemical evolution of the s-process from AGB stars}
\author{\parbox{\textwidth}{\flushleft
\vspace{-0.5cm}
{\it Alessandra Serminato\affil{A}, Roberto Gallino\affil{A}, Claudia Travaglio\affil{C}, Sara Bisterzo\affil{A} and Oscar Straniero\affil{D}}\\
\vspace{0.4cm}
{\small \affil{A}\,Dipartimento di Fisica Generale, Universit\`a di Torino, via P. Giuria, 1, 10125 Torino, Italy}\\
{\small \affil{C}\,Osservatorio Astronomico di Torino (INAF), Strada Osservatorio 20, 10025 Pino Torinese (To),Italy.}\\
{\small \affil{D}\,Osservatorio Astronomico di Collurania (INAF), via M. Maggini, Teramo (64100) Italy}}}
\begin{document}
\twocolumn[
\begin{changemargin}{.8cm}{.5cm}
\begin{minipage}{.9\textwidth}
\vspace{-1cm}
\maketitle
%
%
%%%%%%%%%%%%%     ABSTRACT    %%%%%%%%%%%%%
%Abstract of no more than 200 words here.
\small{\bf Abstract:}
We follow the chemical evolution of the Galaxy for the s-elements
using a Galactic evolutionary model (GCE), as already 
discussed by \cite{Travaglio99,Travaglio01,Travaglio04}, 
with a full updated network and refined asymptotic giant branch (AGB) models. 
%The production of s-process elements in AGB stars proceeds from the 
%combined operation of two neutron sources: the dominant reaction  
%$^{13}$C($\alpha$,n)$^{16}$O, which releases neutrons in radiative 
%conditions during the interpulse phase, and the reaction  
%$^{22}$Ne($\alpha$,n)$^{25}$Mg, marginally activated during 
%thermal instabilities. 
Calculations of the s-contribution to each isotope at the epoch 
of the formation of the solar system is determined by following 
the GCE contribution by AGB stars only.   
Then, using the r-process residual method we determine
for each isotope their solar system r-process fraction,
and recalculate the GCE contribution 
of heavy  elements accounting for both the s- and the r-process.
We compare our results with 
spectroscopic abundances at various metallicities of {[}Sr,Y,Zr/Fe{]}, 
of {[}Ba,La/Fe{]}, of {[}Pb/Fe{]}, 
typical of the three s-process peaks, as well as of {[}Eu/Fe{]},
which in turn is a typical r-process element. Analysis of the various
uncertainties involved in these calculations are discussed.
%Due to the relatively long lifetime of low mass AGBs,
%at lower metallicities the primary r-process 
%contribution plays the dominant role for all heavy elements.\\
%%%%%%%%%%%%%     KEYWORDS    %%%%%%%%%%%%%

\medskip{\bf Keywords:}s and r process --- nuclear reaction --- nucleosynthesis --- abundances ---stars:AGB
% Please write all keywords in lower case. PASA uses the
% standard list of subject headings adopted by The Astrophysical Journal
% and available from
% http://www.journals.uchicago.edu/ApJ/keywords_text.html.
% Keywords are separated by em-dashes, i.e. ---
%%%%%%%%DO NOT EDIT%%%%%%%%%%%%
\medskip
\medskip
\end{minipage}
\end{changemargin}
]
\small
%%%%%%%%EDIT FROM HERE%%%%%%%%%%%%
\section{Introduction}
%metti fig del Ba138 per la massa 1p5
\begin{figure}[h]
\begin{center}
\includegraphics[scale=0.3,angle=270]{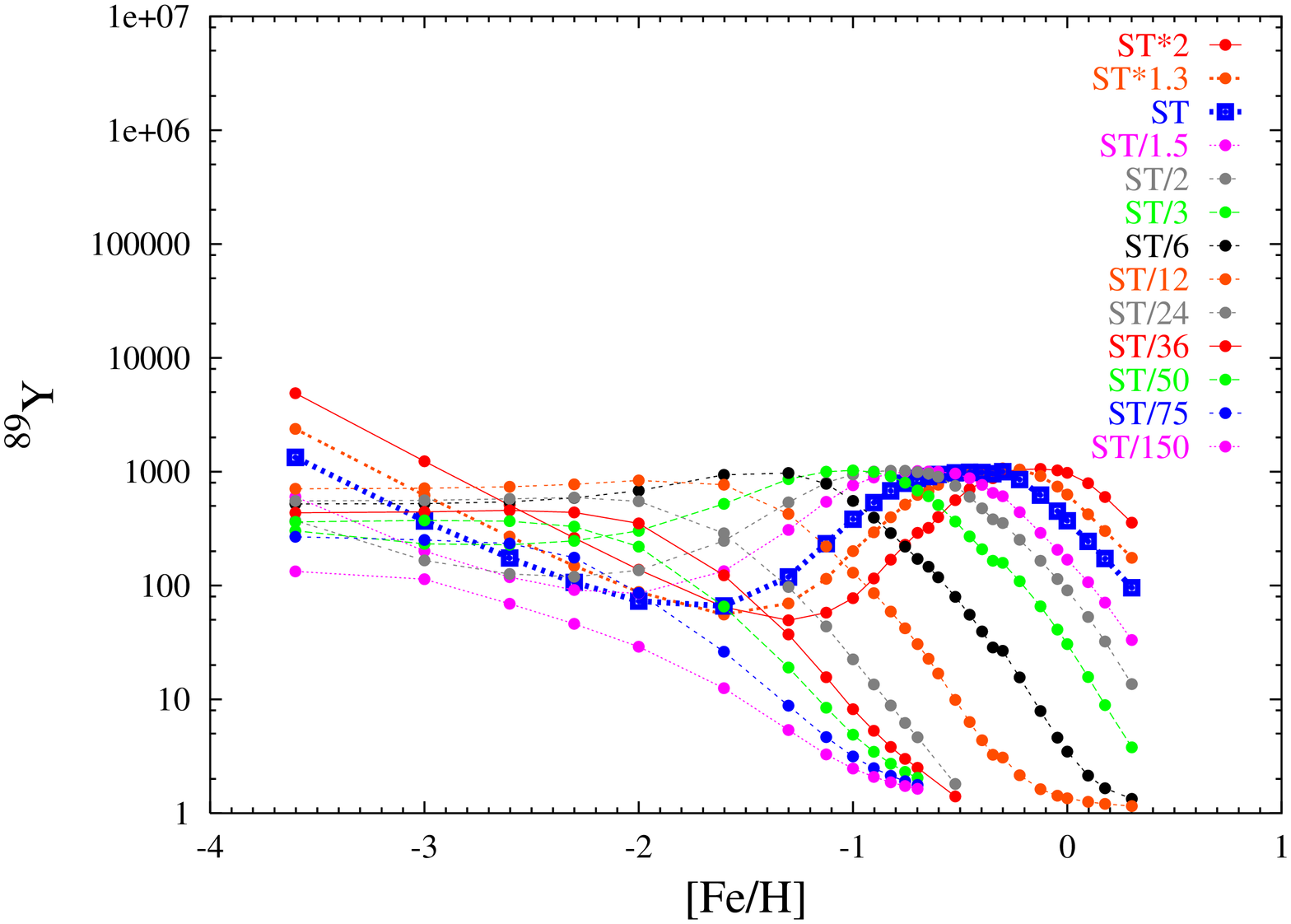}
\includegraphics[scale=0.3,angle=270]{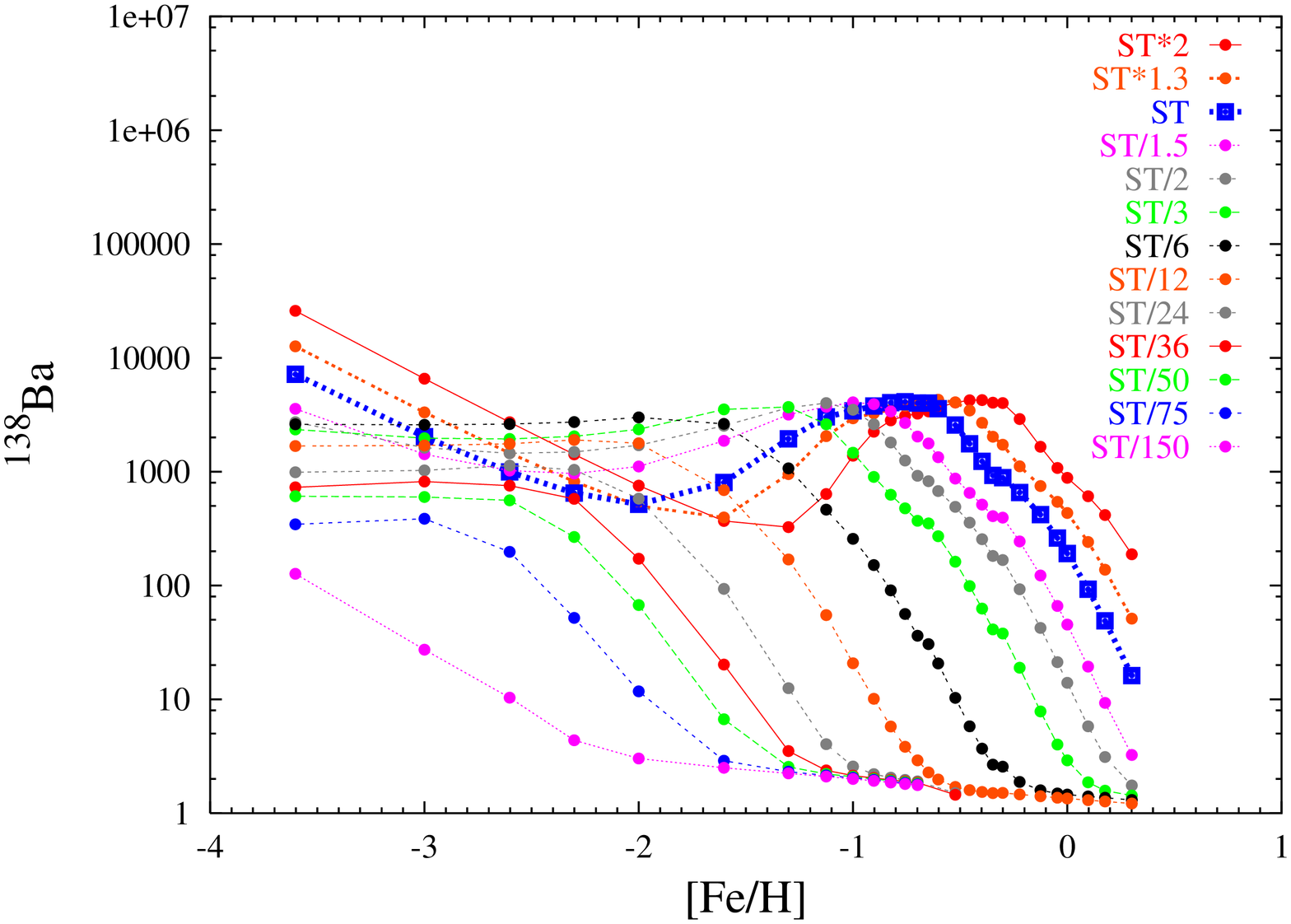}
\includegraphics[scale=0.3,angle=270]{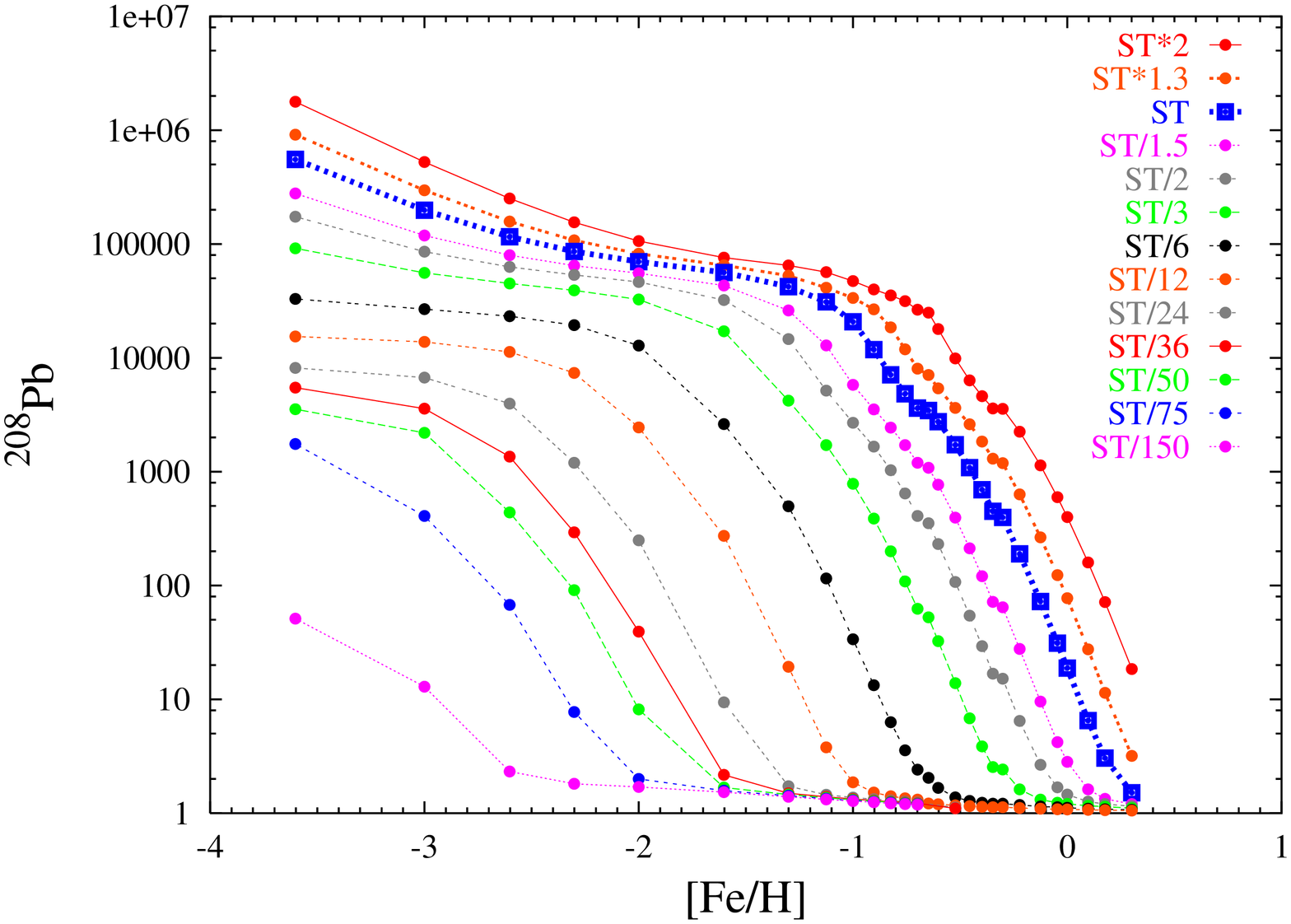}
\caption{Top panel: Theoretical prediction for $^{89}$Y production factors
versus metallicity using AGB models with initial mass $M$ = 1.5 $M_\odot$.
Middle and bottom panel: Analogous plots for $^{138}$Ba and for $^{208}$Pb}
\label{Ba138m1p5}
\end{center}
\end{figure}

\begin{figure}[h]
\begin{center}
\includegraphics[scale=0.3,angle=0]{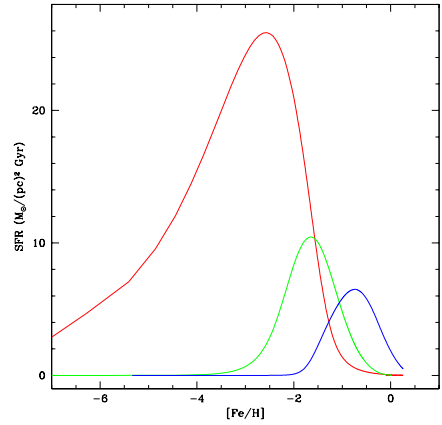}
\caption{Star formation rate versus metallicity.}\label{sfr}
\end{center}
\end{figure}

%metter s: sr y zr ba pb
\begin{figure}[h] 
\begin{center} 
\includegraphics[scale=0.3,angle=0]{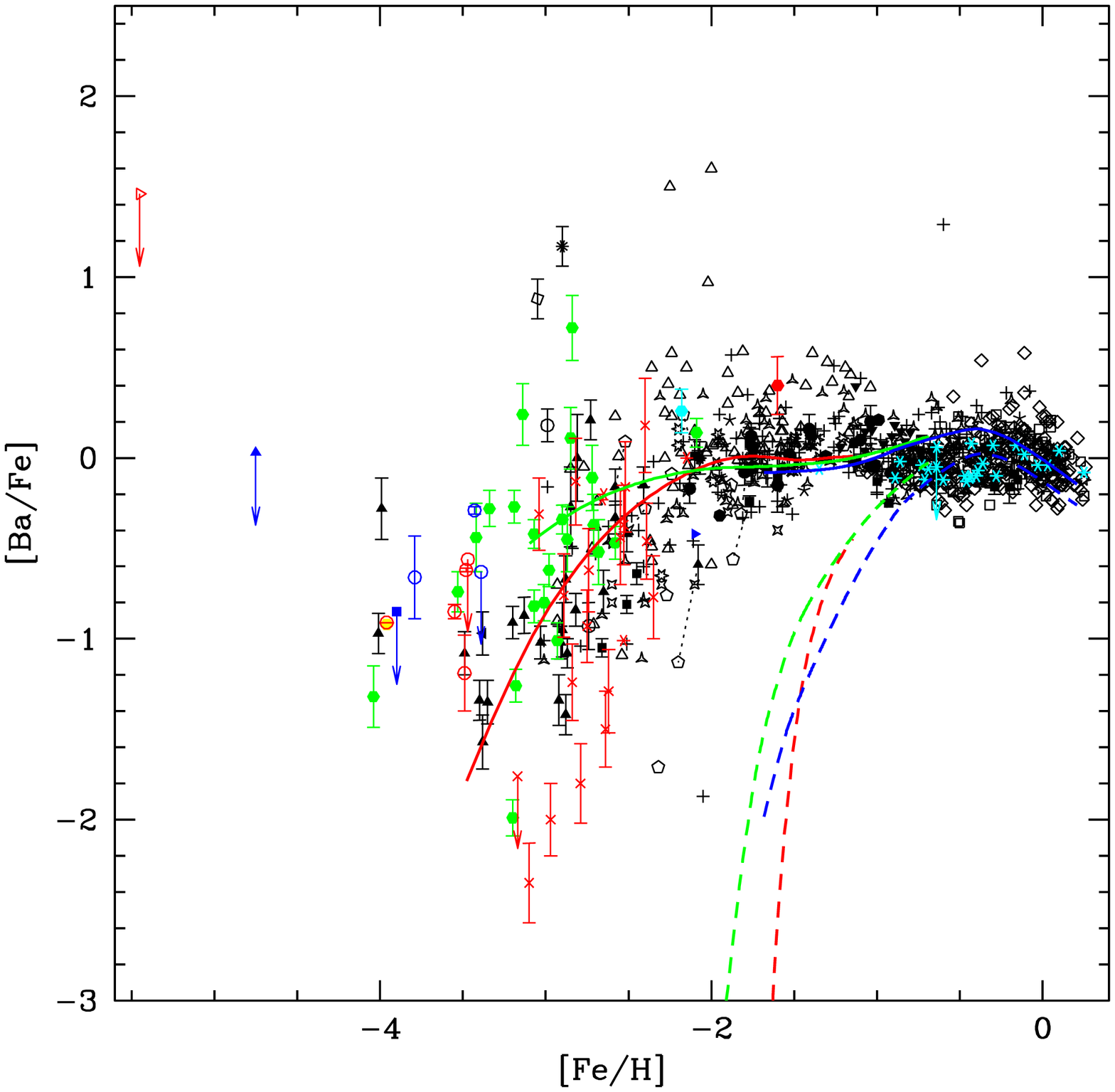}
\includegraphics[scale=0.3,angle=0]{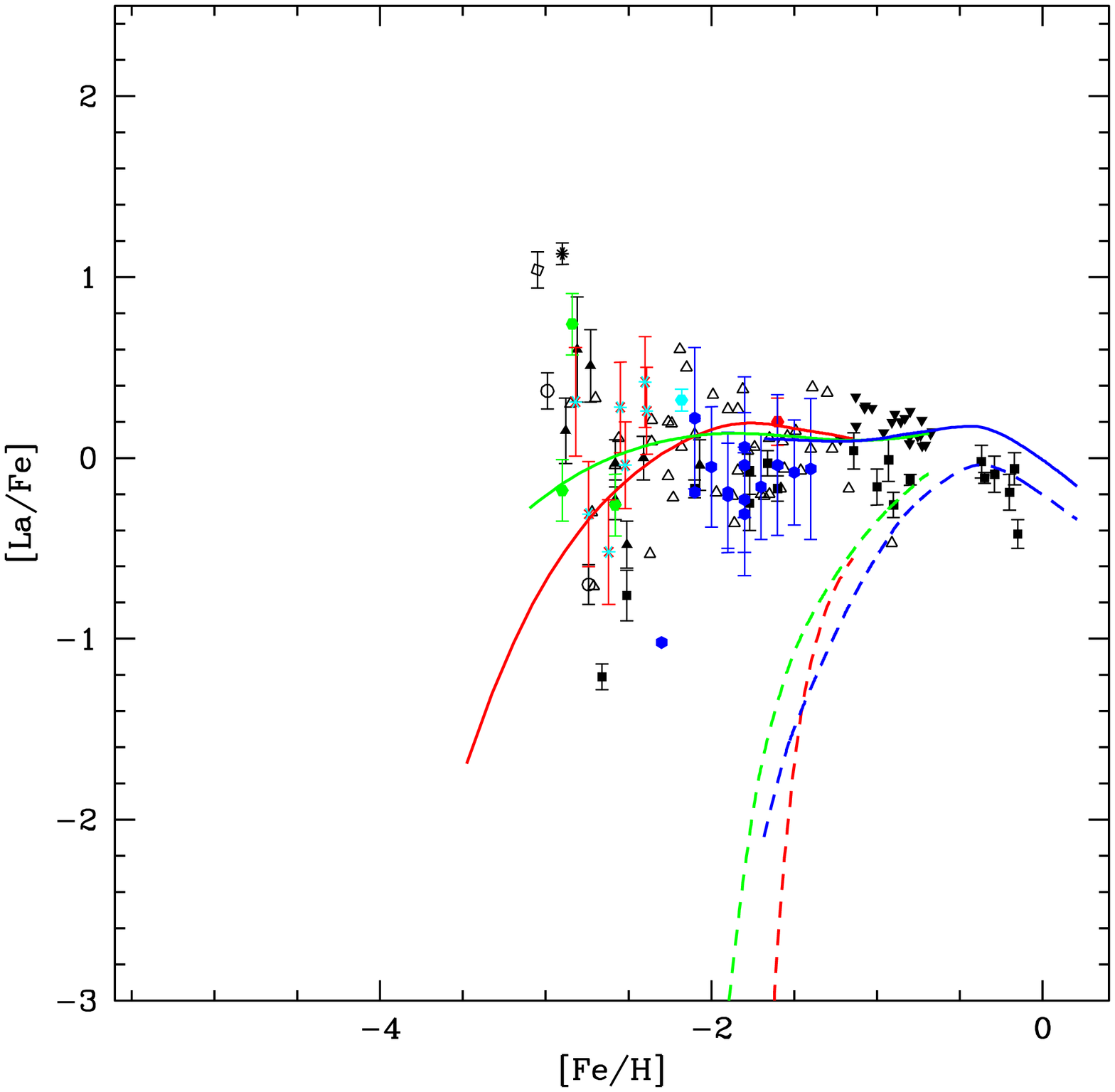}
\caption{Top panel: Evolution of [Ba/Fe] s-fraction as function of [Fe/H]
in the halo, thick disc and thin disc are shown as dashed lines.
Solid lines are for the total s+r Ba theoretical expectations.
Spectroscopic observations of
Galactic disc and halo stars for [Ba/Fe] versus [Fe/H] from literature
(\cite{Travaglio99} implemented with more recent observations as detailed 
in the text). 
Error bars are shown only when  
reported for single objects by the authors. The dotted line connects
a star observed by different authors.
Bottom panel: Analogous plot for [La/Fe].
}\label{bafeagbprova3} 
\end{center}
\end{figure}

\begin{figure}[h]
\begin{center}
\includegraphics[scale=0.3,angle=0]{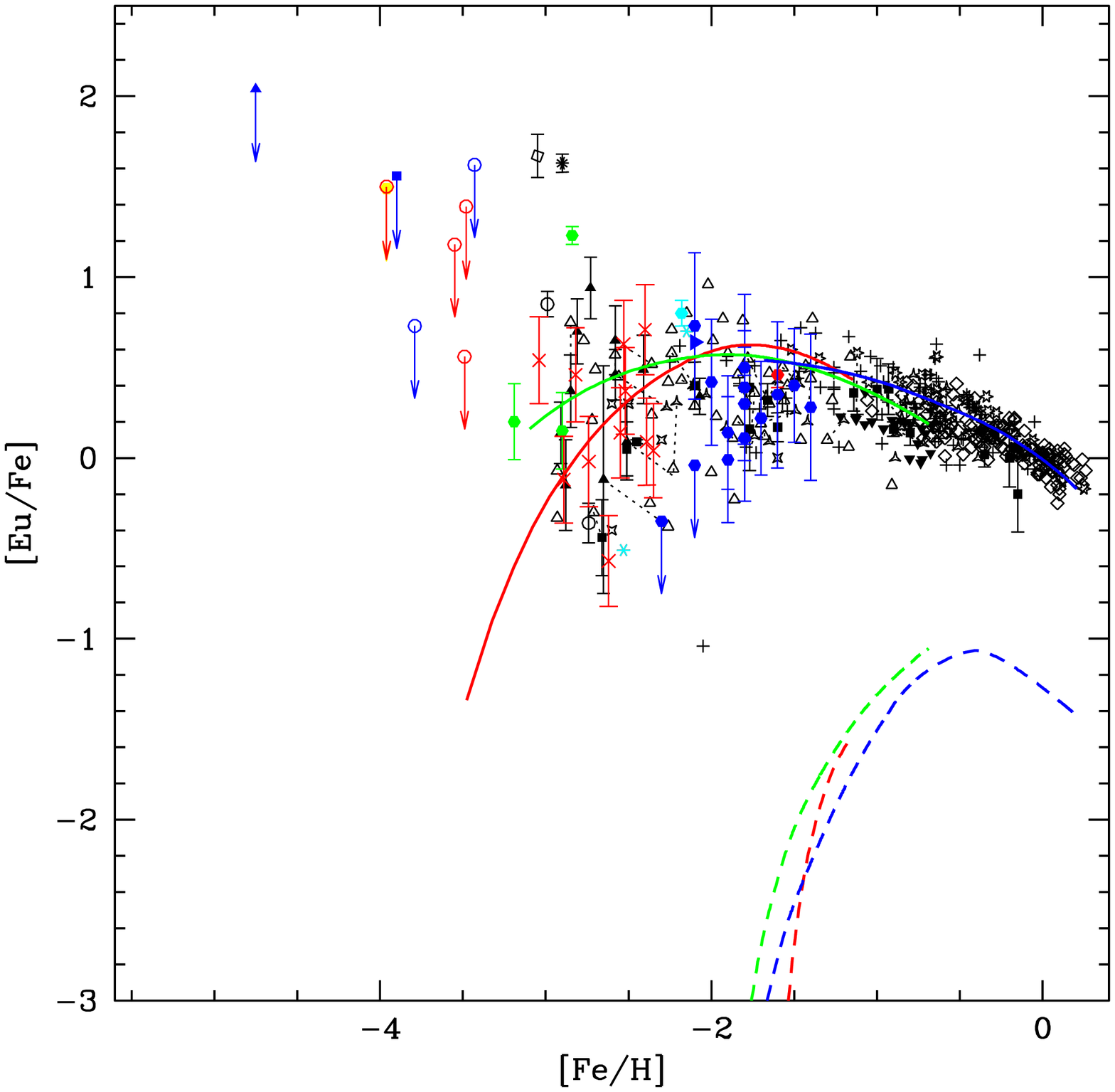}
\includegraphics[scale=0.3,angle=0]{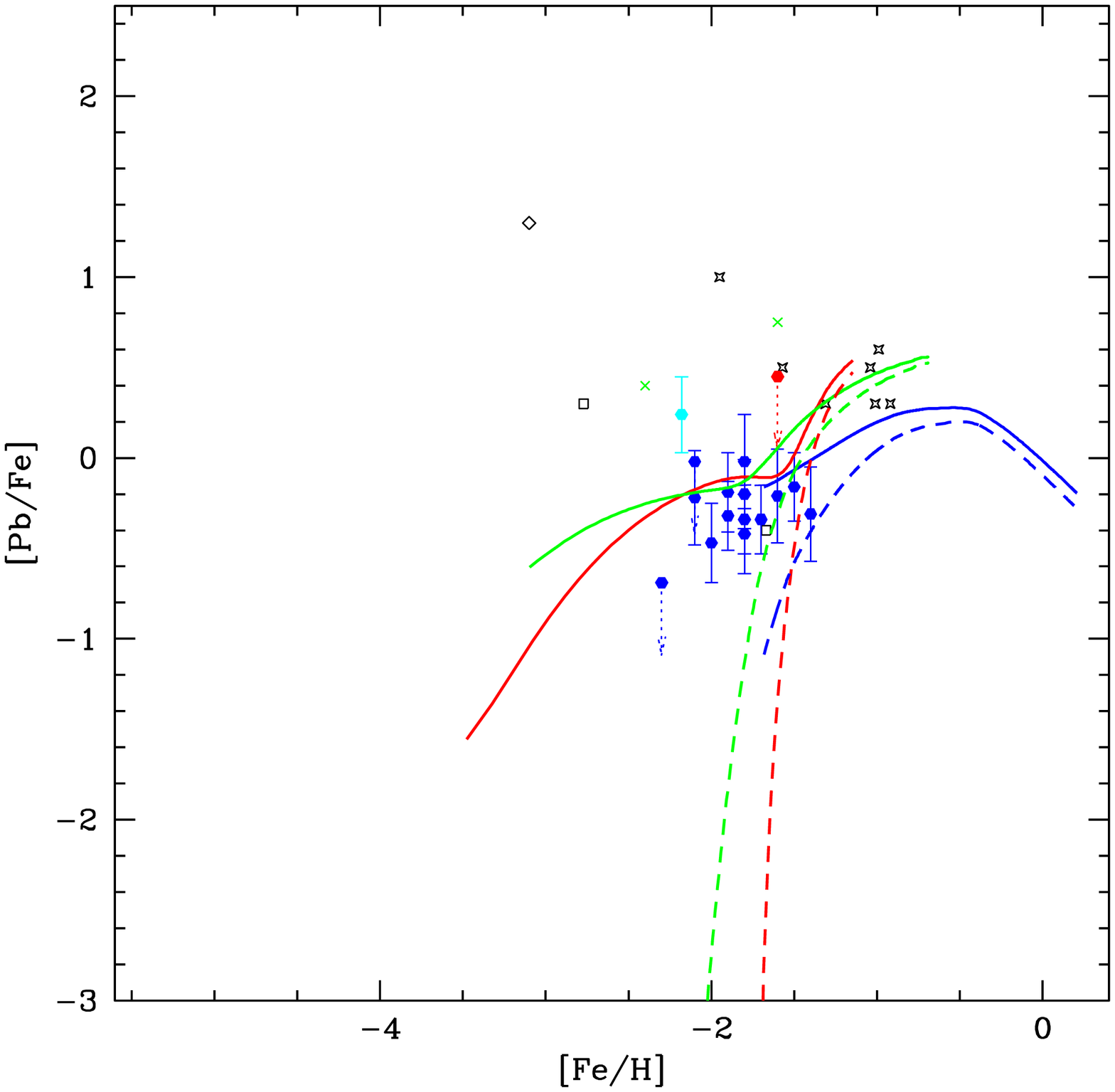}
\caption{Top panel: Galactic chemical evolution of [Eu/Fe] versus [Fe/H]
compared with spectroscopic observations.
Bottom panel: Analogous plot for [Pb/Fe] versus [Fe/H].
}\label{eufetotprova3}
\end{center}
\end{figure}

\begin{figure}[h] 
\begin{center}    
\includegraphics[scale=0.3,angle=0]{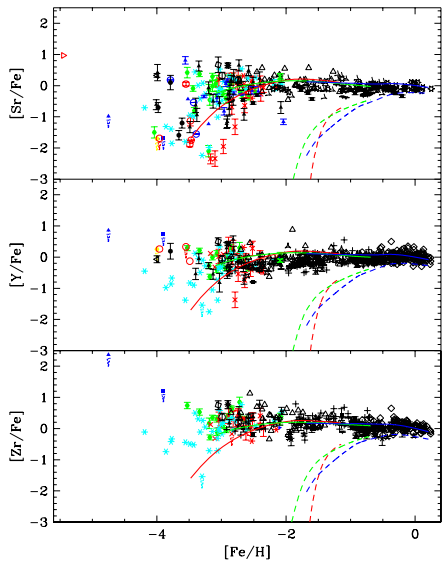}
\caption{Galactic chemical evolution of [Sr/Fe] versus [Fe/H] (upper panel),
[Sr/Fe] versus [Fe/H] (middle panel), and [Zr/Fe] versus [Fe/H] (lower
panel) compared with spectroscopic observations.
}\label{sryzrfeagbprova3}
\end{center}
\end{figure}

\begin{figure}[h]
\begin{center}
\includegraphics[scale=0.3,angle=0]{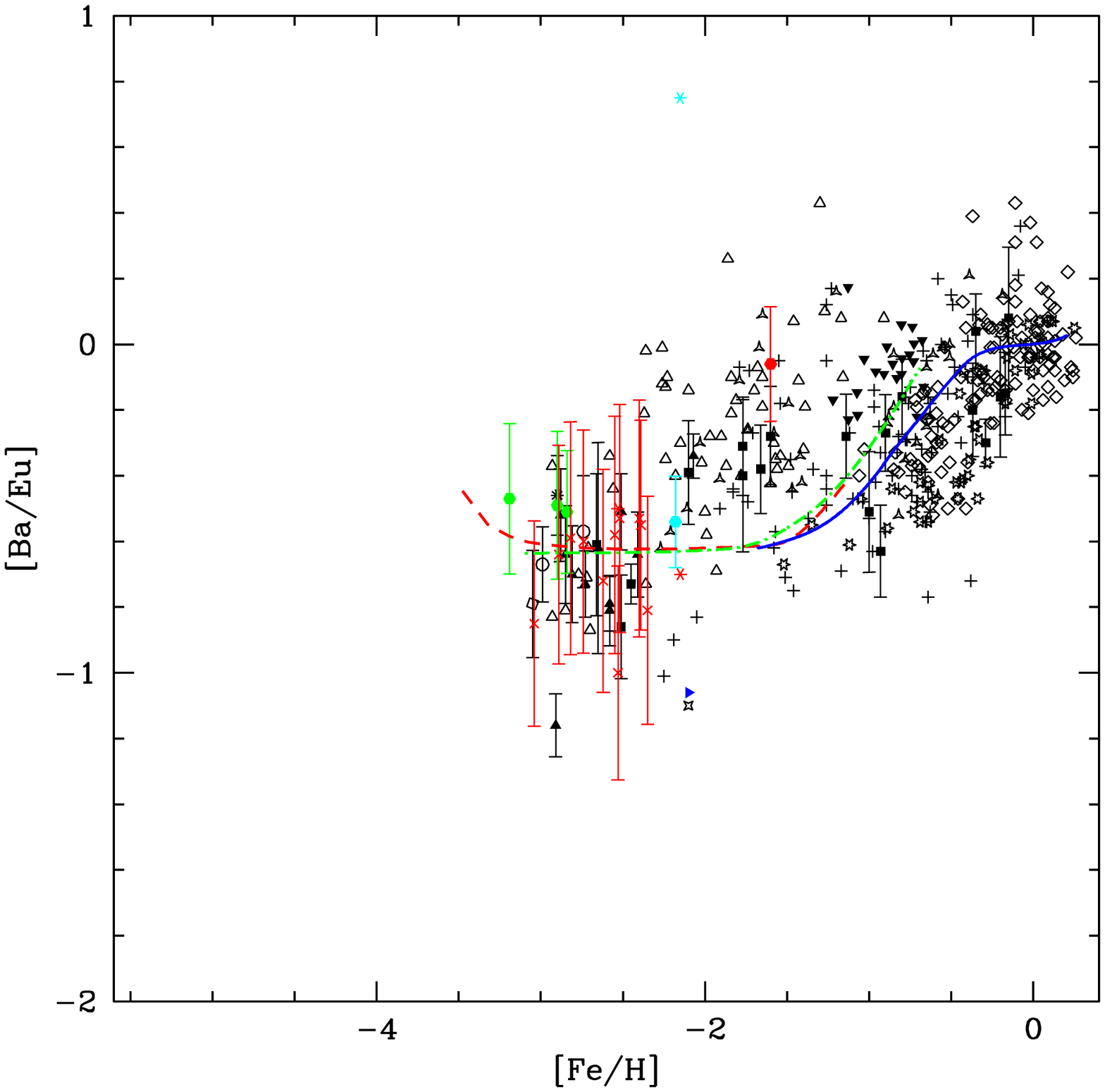}
\includegraphics[scale=0.3,angle=0]{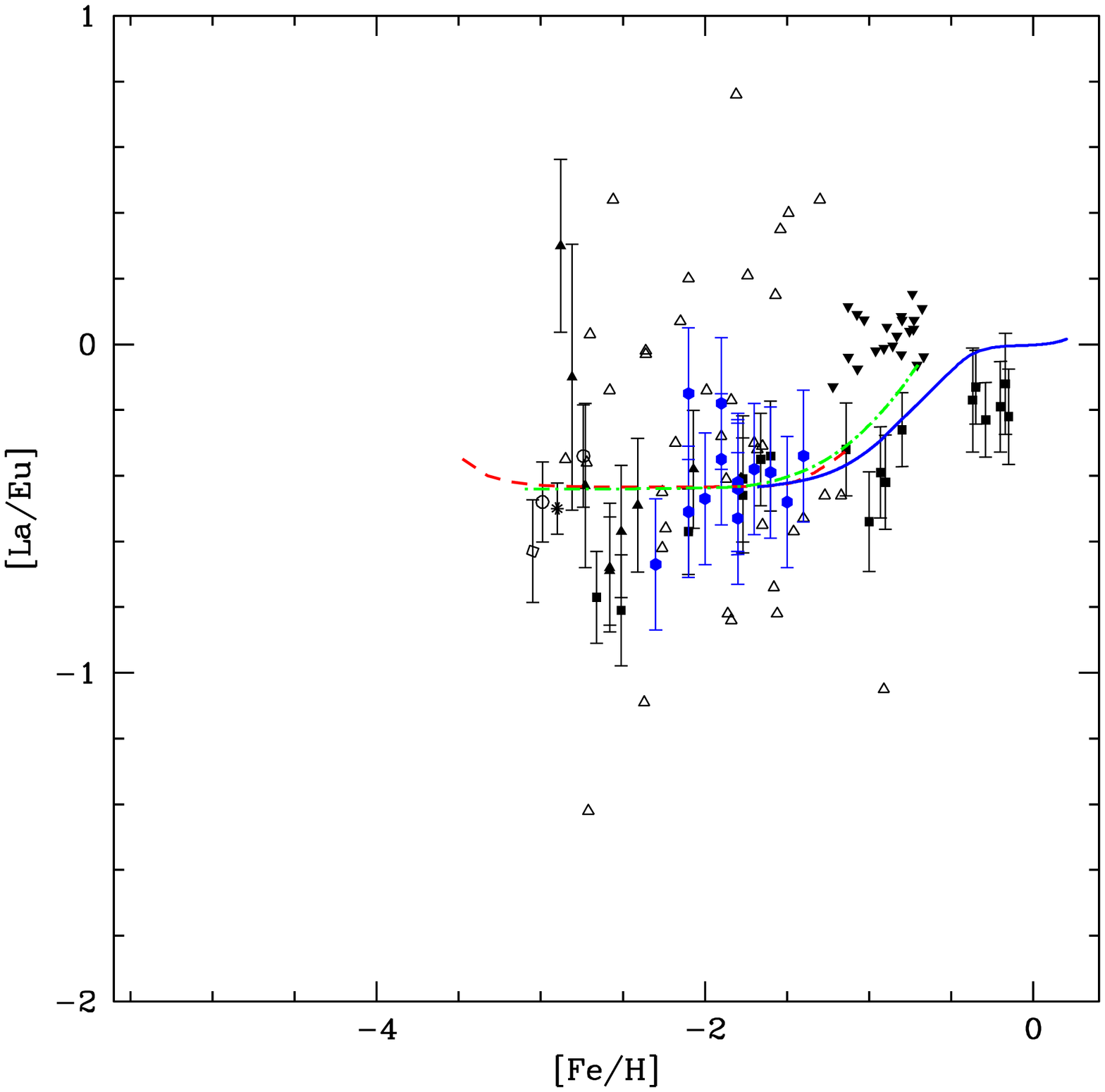}
\caption{Top panel: Galactic chemical evolution of [Ba/Eu] versus [Fe/H]
including both s- and r- process contributions in the thin disc
(long-dashed line), thick disc (dotted line) and halo (solid line).
Error bars are shown only when reported by the
authors. 
Bottom panel: Analogous plot for [La/Eu] versus [Fe/H].
}\label{baeufetotprova3}
\end{center}
\end{figure}

According to the classical analysis of the s-process,
the abundance distribution in the solar system was 
early recognized as the combination of three components (K\"appeler et
al. 1982, Clayton and Rassbach 1977): 
the main component, accounting for s-process isotopes in the range
from  A $\sim$ 90 to  A $<$ 208,
the weak component, accounting  for s-process isotopes up to  A $\sim$ 90, 
and the strong component, introduced to reproduce about 50\% of double magic  
$^{208}$Pb.
The main component itself cannot be interpreted as the result of a
single neutron exposure, but as a multi-component, like 
an exponential distribution of 
neutron exposures. It is clear that the s-process is not originated in a unique
astrophysical environment. 

In this paper we study the Galactic chemical evolution of the s-process as
the outcome of the nucleosynthesis occurring in low to intermediate mass
asymptotic giant branch (AGB) stars of various metallicities.
 These calculations have been performed with an updated network of
neutron capture cross sections and $\beta$ decay rates.
The paper is organized as follows: \S~2 we briefly introduce the stellar
evolutionary model FRANEC  
and the post-process network we  use to compute the nucleosynthesis in AGB 
stars. In \S~3 we introduce the Galactic chemical 
evolution model (GCE) adopted. 
In \S~4 we present the s-elements contributions at the solar system formation 
by introducing in the GCE code the AGB s-yields only obtained at various
metallicities. The corresponding r-process contribution to solar abundances are then
deduced with the r-process residual method.
We recalculate with the GCE model the global s+r contribution to
the Galactic chemical evolution of heavy elements as a function of [Fe/H].
Our predictions are compared with spectroscopic 
data of Sr, Y, Zr, characterising the first s-peak
(light-s, ls), of Ba and La, characterising
the second s-peak (heavy-s, hs), and Pb at the third s-process peak,
together with Eu, an element of most r-process origin.
Finally, in \S~5 we summarise the main 
conclusions and point out few aspects deserving further analysis.

\section{FRANEC and s yields}

The FRANEC (Frascati Raphson Newton Evolutionary code, Chieffi \& Straniero
1999) self-consistently reproduce the third dredge up episodes 
in AGB stars and the 
consequent recurrent mixing of freshly synthesised s-processed material 
(together with $^{4}$He and $^{12}$C) with the surface of the star.
Nucleosynthesis in AGB stars of different masses and metallicities 
is followed with a post-process 
code, which uses the pulse by pulse results of the FRANEC code: 
the mass of the He intershell, the mass involved in the third dredge up
(TDU), the 
envelope mass that is progressively lost by intense stellar winds,
the temporal behaviour of the 
temperature and density in the various layers of the zones 
where nucleosynthesis takes place. For numerical details on the key parameters
affecting the s-process nucleosynthesis in AGB stars of low mass we refer to
Straniero et al. (2003).

The network contains more than 
400 isotopes and is sufficiently extended to take into account 
all possible branchings that play a role in the nucleosynthesis 
process.
The neutron capture network is updated with the recommended (n,$\gamma$) rates by
\cite{Bao00}, complemented by a series of more recent experimental results 
(for more details see Bisterzo et al. 2006).
Stellar $\beta$-decays are treated following Takahashi and Yokoi (1987). 
The production of s-process elements in AGB stars proceeds from the
combined operation of two neutron sources: the dominant reaction
$^{13}$C($\alpha$,n)$^{16}$O, which releases neutrons in radiative
conditions during the interpulse phase, and the reaction
$^{22}$Ne($\alpha$,n)$^{25}$Mg, marginally activated during
thermal instabilities. In the model, the dominant neutron source is not
based on physical principles (Gallino et al. 1998):
during TDU, a small amount of hydrogen from the envelope
may penetrate into $^{12}$C-rich and $^{4}$He-rich (He-intershell) inner zone. 
Then, at H-shell reignition, a thin
$^{13}$C-pocket may form in the top layers of the He-intershell,
by proton capture on the abundant $^{12}$C.
We artificially introduce a $^{13}$C pocket, which is treated as a free
parameter.
The total mass of the $^{13}$C pocket is kept constant with pulse number
and the concentration of $^{13}$C in the pocket is varied in a large range,
from values 0.005 $-$ 0.08 up to 2 times with respect to the profile indicated as ST by 
\cite{Gallino98}, corresponding to the choice of the mass of  
$^{13}$C  of 3.1 $\times$ 10$^{-6}$ $M_\odot$. 
A too high proton concentration would favour the production of $^{14}$N
by proton capture on $^{13}$C. Note that the minimum $^{13}$C-pocket efficiency  
decreases with metallicity, since the neutron exposure depends on the ratio of
the neutrons released to Fe seeds.
This choice was shown to better reproduce the main component with AGB models
of half-solar metallicity (Arlandini et al. 1999), and is a first
approach to the understanding of solar system s-process abundances.
In reality, the solar system composition is the outcome of all previous
generations of AGB stars having polluted the interstellar medium up to
the moment of condensation of the solar system.
A spread of $^{13}$C-pocket efficiencies has been shown
to reproduce  observations of s-enhanced stars at
different metallicities (see, e.g., Busso et al. 1999, 2001, Sneden et al.
2008).

In AGB stars of intermediate mass the s-process is less efficient.
As for the choice of the
$^{13}$C neutron source, because of the much shorter interpulse
phases in these stars ($\sim$6500 yr for 5 $M_\odot$ and
$\sim$1500 yr for 7 $M_\odot$) with respect to LMS-AGBs
($\sim$3$-$6 $\times$ 10$^{4}$ yr), the He intershell mass involved
is smaller by one order of magnitude. Consequently, also the TDU of
s-process-rich material from the He-intershell into the surface is
reduced, again by roughly one order of magnitude. Given the above
reasons, for the 5 $M_\odot$ and the 7 $M_\odot$ cases, as in
Travaglio et al. (1999, 2004), we have considered as a standard choice for
IMS-AGBs (ST-IMS) a $^{13}$C mass scaled accordingly
[$M$($^{13}$C)$_{\rm ST-IMS}$= 10$^{-7}$ $M_\odot$].
On the other hand, in IMS stars the 
$^{22}$Ne($\alpha$,n)$^{25}$Mg reaction is activated more 
efficiently (Iben 1975; Truran \& Iben 1977) since the temperature 
at the base of the convective pulse reaches values of 
$T$ = 3.5$\times$10$^{8}$ K. Also, the 
peak neutron density during the TP phase is consistently higher 
than in  AGBs ($N_n \sim$ 10 $^{11}$ n cm$^{-3}$, see 
Vaglio et al. 1999; Straniero et al. 2001), overfeeding a few neutron-rich 
isotopes involved in important branchings along the s-process path, 
such as $^{86}$Kr, $^{87}$Rb and $^{96}$Zr.

We took a set of low mass stars (LMS) (1.5 and 3 solar masses) and 
intermediate mass stars (IMS) (5 and 7 solar masses), and a set of 
27 metallicities from [Fe/H] = 0.30 down to [Fe/H] = $-$3.60.

\subsection{s-yields}

In Fig. \ref{Ba138m1p5} we show the theoretical predictions versus [Fe/H], for AGB 
stars of initial mass $M$ = 1.5 $M_\odot$, of the production factors
in the astrated s-process ejecta of $^{89}$Y, $^{138}$Ba and  $^{208}$Pb, taken as 
representative of the three s-process peaks. Each line corresponds to a given
$^{13}$C-pocket efficiency. The production factors are given
in terms of the isotope abundance  divided
by the initial abundance, solar-scaled with metallicity.
For low neutrons/seed ratios, the neutron fluence mainly feeds the ls nuclei
(like $^{89}$Y), whereas for higher exposures the 
hs peak  (like $^{138}$Ba) is favoured. 
Increasing further the neutron exposure, the neutron flow tends to overcome the
first two s-peaks, directly feeding $^{208}$Pb at the termination point of
the s-proces path.
There is therefore a very complex s-process dependence 
on metallicity. 

%____________________________________________________________________

\section{Galactic chemical evolution model}
%metto una fig della SFR
% It is preferable to embed your figures in the text as in the 
% following example
%\begin{figure}[ht]
%\begin{center}
%\includegraphics[scale=0.3,angle=0]{serminatopasaSFRfig3.eps}
%\caption{Star formation rate vs metallicity.}\label{sfr}
%\end{center}  
%\end{figure}

The model for the chemical evolution of the Galaxy was described in 
detail by \cite{Ferrini92} and it was updated by 
\cite{Travaglio99,Travaglio01,Travaglio04}. 
The Galaxy is divided into three zones, halo, thick disc and thin 
disc, whose composition of stars, gas (atomic and molecular) and 
stellar remnants is computed as function of time up to the present 
epoch $t_{\rm Gal}$ = 13 Gyr. Stars are born with an initial 
composition equal to the composition of the gas from which they 
formed. The formation of the Sun takes place 4.5 Gyr ago, at epoch 
$t_\odot$ = 8.5 Gyr.
The matter in the Galactic system has different phases of 
aggregation, interacting and interchanging one into the other. 
Therefore the evolution of the system (the time dependence of the 
total mass fraction in each phase and the chemical abundances in the 
ISM and in stars) is determined by the interaction between these 
phases. 
It means that the star formation rate (SFR) 
(see Fig. \ref{sfr}) $\psi(t)$ is not assumed $a priori$, but is 
obtained as the result of a self-regulating process occurring in the 
molecular gas phase, either spontaneous or simulated by the 
presence of other stars. The thin disc is divided into concentric 
annuli, without any radial flow, and is formed from material 
infalling from the thick and the halo. In the present work, as in 
Travaglio et al. previous works, we neglect any dependence on 
Galactocentric radius in the model results as well as in the 
observational data and we concentrate on the evolution inside the 
solar annulus, located at 8.5 kpc from the Galactic center.

However, we must point out that the Galactic chemical evolution
model by \cite{Ferrini92} that we use is now believed to be incorrect. 
The main problem is that the thick disk cannot form from gas from the
halo, as demonstrated by \cite{WG1992}. These authors showed that the
distribution of angular momentum of halo stars differs markedly from
that of the thick and thin disks. \cite{Pardi1995} also demonstrated
that the scenario we assume cannot reproduce at the same time the
stellar metallicity distributions of the halo, thick disk, and thin
disk. See also \cite{Pagelbook} and \cite{Matteuccibook}.  To overcome
the problems of the model by \cite{Ferrini92}, \cite{Cescutti2006}
studied the chemical evolution of the heavy elements using the
two-infall model proposed by \cite{Chiappini1997}.  Also this model,
although widely adopted, presents some shortcomings, for example,
it is not possible to distinguish the thick disk from the thin 
disk. In the present paper we focus on analysing the changes made 
by using updated reaction rates on the chemical evolution of the 
elements heavier than iron rather than the changes made by using 
an updated model of the evolution of the Galaxy. Thus, we use 
the same model of \cite{Travaglio99,Travaglio01,Travaglio04} where 
we introduce a new and extended grid of AGB yields.

\section{Results for the Galactic chemical evolution of s- and r- elements}

In this section we present the results for the evolution of  
Sr, Y, Zr, La, Ba, Eu and Pb in the Galaxy, by considering 
separately the s- and r-contributions. 
Then we compute the Galactic abundances of these elements resulting from the 
sum of the two processes, comparing model results with the available 
spectroscopic observations of field stars at different metallicities.
 
\subsection{Galactic chemical evolution of s-elements}

The s-contribution to each isotope at the epoch
of the formation of the solar system is determined by following
the GCE heavy elements contributed by AGB stars only.
Then, using the r-process residual method (s = 1 $-$ r) we determined
for each isotope the solar system r-process fraction.
As a second step, we recalculate the GCE contribution
of the heavy elements accounting for both the s- and the r-process,
assuming that the production of r-nuclei is a primary
process occurring in Type II supernovae, independent of the
metallicity.

Galactic chemical s-process expectations depend on several uncertainties,
among which are the knowledge of solar abundances, of the neutron capture
network and on the choice of the specific stellar evolutionary code.   
To this one may add the uncertainties connected with the treatment of the 
Galactic chemical evolution model.
Among the most important uncertainties is the evaluation of the global
ejecta from the AGB winds of stars of different masses and
metallicities, which in turn depend on the mass mixed with the
envelope by the various third dredge up episodes, and by the  
the weighted average s-process yields over the assumed $^{13}$C-pocket efficiencies.   
This would provide a very poor expectation.
However, a strong constraint is given by the heavy s-only isotopes, whose
solar abundance derives entirely from the s-process in AGB stars. Among the s-only
isotopes, the unbranched $^{150}$Sm, whose neutron capture cross section at
astrophysical temperatures and solar abundance are very well known, with
a total uncertainty of less than 3\% (Arlandini et al. 1999), may be chosen
as normalisation. One may then deduce the relative
s-process isotope percentage for all heavy elements.

For LMS we averaged the s-process yields over 13 $^{13}$C-pocket,
excluding the case ST $\times$ 2.
For IMS, the effect of the $^{13}$C neutron source is negligible with
respect to the one induced by $^{22}$Ne neutron source. 

In Table \ref{comparisonlmsnorm}  
we show values of AGB percentage to solar abundance at $t$ = $t_\odot$
for LMS and IMS respectively obtained by present calculations
compared with  \cite{Travaglio99} results. In Table 1 a choice of selected
isotopes is made, among which the s-only isotopes $^{124}$Te, $^{136}$Ba,
$^{150}$Sm and $^{204}$Pb, together with $^{89}$Y, $^{138}$Ba, and
$^{208}$Pb of major s-process contribution.  In turn $^{151}$Eu is chosen
as representative of the r-process, as clearly indicated by its only 6\%
to solar $^{151}$Eu.

We compare our results with
spectroscopic abundances of [Sr,Y,Zr/Fe], [Ba,La/Fe], and [Pb/Fe] that are
typical of
the s-process peaks, as well as [Eu/Fe],
which in turn is a typical r-process element.

Let us first consider [Ba/Fe] and [La/Fe] versus [Fe/H].
Figs. \ref{bafeagbprova3} show in the top panel the
[Ba/Fe] versus [Fe/H] with spectroscopic observations and
theoretical s-curves, and in the bottom panel the analogous plot
[La/Fe] versus [Fe/H]. 
In this figure and the following we compare with the set of 
stellar observations used by \cite{Travaglio99,Travaglio01,Travaglio04}
implemented with more recent observations of elemental abundances 
in field stars, as listed below, with their associated symbols in the figures:
\cite{Mashonkina06} blue asterisks; 
\cite{Ivans06} cyan full hexagons;
\cite{AokiB08} red open squares; 
\cite{AokiH08} blue asterisks; 
\cite{Lai08} green full hexagons; 
\cite{Cohen07} yellow full hexagons;
\cite{Norris07} blue full triangles; 
\cite{Frebel07} full blue squares; 
\cite{Mashonkina08} red asterisks; 
\cite{Roederer08} full red hexagons; 
\cite{AokiH05} red crosses; 
\cite{Francois07} cyan asterisks; 
\cite{Cohen08} red open circles; 
\cite{AokiF06} red open triangles pointing to the right; 
\cite{Yushch05} blue full triangles; 
\cite{Vaneck} black open triangles; 
\cite{Cowan02} green crosses. 
 The dashed lines show the theoretical GCE expectations 
using only the AGB s-process products for halo, thick and thin disc
separately. Although the s-contributions to solar Ba and La are 78.2\% and
66.3\%, respectively, it is clear that s-process alone does not
explain all spectroscopic observations.

In Fig. \ref{eufetotprova3} analogous plots are shown for [Eu/Fe] (top
panel) and [Pb/Fe] (bottom panel) versus [Fe/H]. While the s-process
contribution to Eu is negligible (5.6\% to solar Eu), the s-contribution to solar Pb is 
83.9\%. Comparing with previous plots, spectroscopic [Pb/Fe] observations
are scarce because of the difficulty to extract Pb abundances from unevolved
stars.
 As we explained
before, the classical analysis of the main component cannot
explain the $^{208}$Pb abundances. The GCE calculation provide
83.9\% to solar Pb, and 91.1\% to $^{208}$Pb, thanks to the contribution of
different generations of AGB stars. In particular,
low metallicities AGB stars are the main contributors to $^{208}$Pb.

Finally, in Fig. \ref{sryzrfeagbprova3} are presented the
analogous plots for [Sr/Fe] versus [Fe/H] (top panel), [Y/Fe] versus [Fe/H] 
(middle panel), and [Zr/Fe] versus [Fe/H] (lower panel).
GCE calculations provide 64.1\% to solar Sr, 66.5\% to solar Y, and 60.3\%
to solar Zr. Note that the classical analysis of the main component would
provide 85\%,  92\%, and  83\%, respectively (Arlandini et al.
1999), making clear also in this case that the classical analysis is only a 
rough approximation.

\begin{table}[h]
\begin{center}
\caption{Galactic LMS-AGB (1.5 to 3 $M_\odot$) 
 and IMS-AGB (5 to 8 $M_\odot$) contributions, at $t$ = $t_\odot$ = 8.5Gyr,
expressed as percentages to solar abundances.}\label{comparisonlmsnorm} 
\begin{tabular}{lcc}
\\
\hline
\hline isotope & Travaglio 99 & our work \\
\hline
   &  LMS-AGB (\% to solar)&   \\
\hline
 $^{89}$Y       &    61.5      &    62.7           \\
 $^{124}$Te     &    72.0      &    70.0           \\
 $^{136}$Ba     &    92.1      &    85.1           \\
 $^{138}$Ba     &    84.0      &    82.3           \\
 $^{139}$La     &    61.4      &    65.5           \\
 $^{150}$Sm     &    98.1      &    99.1          \\
 $^{151}$Eu     &     6.4     &    5.7           \\
 $^{204}$Pb     &    93.8      &    85.1           \\
 $^{208}$Pb     &    93.6      &    90.7
          \\
 \\
 \hline
 \hline    &  IMS-AGB (\% to solar)&   \\
\hline
$^{89}$Y       &     7.5     &    3.8             \\
$^{124}$Te     &     4.7     &    2.2             \\
$^{136}$Ba     &    4.1      &   2.2          \\
$^{138}$Ba     &    2.5      &   1.2           \\
$^{139}$La     &    1.7      &   0.8           \\
$^{150}$Sm     &     2.8     &    0.9           \\
$^{151}$Eu     &     0.2     &    0.06           \\ 
$^{204}$Pb     &     2.5     &    0.9            \\
$^{208}$Pb     &     1.2     &    0.4             \\
\hline
\end{tabular}
%\medskip\\
%$^a$ Travaglio et al 1999.\\
%$^b$ Our work.\\   
\end{center}
\end{table}

\subsection{The r- process yields and Galactic chemical evolution}

From the theoretical point of view, the r-process origin is
still a matter of debate. The analytical approach followed here to
derive the r-process yields has been presented first
by \cite{Travaglio99}. The enrichment of r-process elements in
the interstellar medium (ISM) during the evolution of the Galaxy
is quantitatively constrained on the basis of the results for the
s-process contribution at $t$ = $t_\odot$.
The so called r-process residual for each isotope is obtained by
subtracting the corresponding s-process contribution
N$_s$/N$_\odot$ from the fractional abundances in the solar system
taken from \cite{AndersGrevesse89}:
\begin{equation}
N_{s}/N_{\odot}= (N_{\odot}-N_{s})/N_{\odot}
\end{equation}
In the case of Ba \cite{Travaglio99} obtain a r-residual of 21$\%$.
The assumption that the r-process is of primary nature and
originates from massive stars allows us to estimate the contribution
of this process during the evolution of the Galaxy. In the case of
Ba, for example
\begin{equation}
(\frac{\rm Ba}{\rm O})_{r,\odot}\sim 0.21(\frac{\rm Ba}{\rm O})_{\odot}.
\end{equation}
Since the s-process does not contribute at low metallicity for Population II
stars
\begin{equation}
\left(\frac{\rm Ba}{\rm O}\right) \sim \left(\frac{\rm Ba}{\rm
O}\right)_{r,\odot}
\end{equation}
assuming a typical [O/Fe]$\sim$0.6 dex for Population II stars. Thus,
the r-process contribution for [Fe/H]$\leq$ $-$1.5 dominates over
the s-contribution and roughly reproduces the observed values.

The procedure followed to extrapolate the r-process yields is 
independent of the chemical evolution model adopted and has been 
described in \S~2.
The solution shown in the plots adopts SNIIe in the mass range 
8 $\leq$ $M$/$M_\odot$ $\leq$ 10 as primary producers of r-nuclei. 
The [element/Fe] ratios
provide information about the enrichment relative to Fe in the
three Galactic zones, making clear that a delay in the r-process
production with respect to Fe is needed in order to match the
spectroscopic data at [Fe/H] $\leq$ $-$2. 
The observations show
that [Ba/Fe] begin to decline in metal-poor stars
and this trend can be naturally explained by the finite lifetimes of
stars at the lower end of the adopted mass range: massive stars in
the early times of evolution of the Galaxy evolve quickly, ending a
s SNII producing O and Fe. Later, less massive stars explode as SNII,
producing r-process elements and causing the sudden increase in
[element/Fe]. At [Fe/H] $\sim$ $-$1 halo stars, thick disc stars and
thin disc stars are mixed up.

The large scatter observed in [Ba/Fe], [La/Fe] and in [Eu/Fe] in halo stars 
can be ascribed to an incomplete mixing in the Galactic halo. 
This allows the formation of very metal-poor stars
strongly enriched in r-process elements, like CS 22892-052
(Sneden 2000a). This star, with [Fe/H] $\sim$ $-$3.1, shows r-process enhancements
of 40 times the solar value ([Eu/Fe] $\sim$ +1.7), and
[Ba/Fe] $\sim$ +0.9. 
Nevertheless its [Ba/Eu] is in agreement with the typical r-process ratios.

\subsection{The s+r process evolution}
The global results for the Galactic 
chemical evolution of heavy elements from iron to lead based on 
the assumptions discussed before, namely that the s-process 
contribution of these elements derives from low mass AGB stars 
and the r-process contribution originates from SNII in the range 
8 $\leq$ $M$/$M_\odot$ $\leq$ 10, are shown as solid lines in Figs. 3, 4, 5.

Fig. \ref{baeufetotprova3} show 
[Ba/Eu] versus [Fe/H] (top panel) and [La/Eu] versus [Fe/H] (bottom
panel) for spectroscopic observations and 
theoretical curves computed by adding the s and r process 
contribution. Since Eu is mostly produced by r-process 
nucleosynthesis (94$\%$ at $t$ = $t_\odot$), the [element/Eu] 
abundance ratios (bottom panel) provide a direct way to judge 
the relative importance of the s and r channels during the evolution 
of the Galaxy. At low metallicity the r-process contribution is 
dominant, and the [element/Eu] ratio is given by the 
elemental r-fraction computed with the r-residuals described before. 
On the other hand, for [Fe/H] $\geq$ $-$1.5, the s-process 
contribution takes over, and the [element/Fe] ratios rapidly 
increase approaching the solar values. 

For elements from Ba to Pb, the estimated r-process contribution at 
$t$ = $t_\odot$ has been derived by subtracting the s-fraction from 
solar abundances (r-process residual method). Instead, for elements lighter 
than Ba a more complex treatment is needed. 
In particular for Sr, Y and Zr, besides the s-process component,
one has to consider three other components: the weak-s component (which
decreases linearly with the metallicity), the r-component and 
the LEPP-component, which are both independent of metallicity (Travaglio et al. 2004).
As reported above, the GCE contribution by AGB stars are 64.1\% to solar Sr, 
66.5\% to solar Y, and 60.3\% to solar Zr. 
The weak s-process is estimated to contribute to 9\% to solar Sr,
10\% to solar Y, and 0\% to solar Zr. 
This leaves for the LEPP component a contribution of 17.9\% to solar Sr,
18.5\% to solar Y, and 28.7\% to solar Zr very close to Travaglio et al.
(2004) expectations. The residual r-process contributions would then be 
9\% of Sr, 5\% of Y and 11\% of Zr.
Summing up all contributions, the solid lines shown in
Fig. 5 give a good explanation of spectroscopic data, both in the halo and
in the Galactic disc.  A refined analysis is difficult to determine and 
is still matter of debate. 

\section{Conclusions}
We have studied the evolution of the heavy elements in the Galaxy, 
adopting a refined set of models for
s-processing in AGB stars of different metallicities and compared 
with observational constraints of unevolved field stars for Sr, Y, Zr, 
Ba, La, Eu and Pb.
In the first part stellar yields for s-process elements have been 
obtained with post-process calculations based on AGB models with 
different masses and metallicities, computed with FRANEC. 

In the second part we have adopted a Galactic chemical evolution model 
in which the Galaxy has been divided into three zones (halo, thick 
disc and thin disc), whose composition of stars, gas (atomic and 
molecular) and stellar remnants, is computed as a function of time 
up to the present epoch. Introducing as a first step in the GCE model the AGB s-yields 
only, we have obtained the s-process enrichment of the Galaxy at the 
time of formation of the solar system. Major uncertainties connected with
the AGB models, with the adopted average of the large spread of $^{13}$C-pocket 
efficiencies, as well as of the basic parameters introduced in the CGE model
are strongly alleviated once we normalise the s-process isotope abundances
computed at the epoch of the solar formation to
$^{150}$Sm, an unbranched s-only isotope with both a well determined solar
abundance and neutron capture cross section at astrophysical temperatures.

Assuming that the production of r-nuclei 
is a primary process occurring in SNII of 8$-$10 solar masses, 
the r- contribution to each nucleus has then been computed as the 
difference between its total solar abundance and its s-process abundance. 
Finally we compare our predictions with spectroscopic observations of the
above listed elements along the life of the Galaxy.

%%Format tables as in the following example
%\begin{table}[h]
%\begin{center}
%\caption{Example Table}\label{tableexample}
%\begin{tabular}{lcc}
%\hline Column 1 & Column 2 & Column 3 \\
%\hline Table Content$^a$ \\
%\hline
%\end{tabular}
%\medskip\\
%$^a$Table footnotes go here.\\
%\end{center}
%\end{table}

\section*{Acknowledgments}
We acknowledge the anonymous referees for their very useful comments.
We thank Maria Lugaro for a very careful reading of the manuscript.
Work supported by the Italian MIUR-PRIN 2006 Project "Final Phases 
of Stellar Evolution, Nucleosynthesis in Supernovae, AGB Stars, 
Planetary Nebulae".\\

%\end{multicols}

\end{document}